\documentclass[longauth]{aa}  

\usepackage{graphicx}
\usepackage[]{hyperref}
\usepackage{txfonts}
\usepackage{xcolor}
\usepackage{dirtytalk}
\usepackage{float}
\usepackage{soul}
\outer\def\gtae {$\buildrel {\lower3pt\hbox{$>$}} \over 
{\lower2pt\hbox{$\sim$}} $}
\outer\def\ltae {$\buildrel {\lower3pt\hbox{$<$}} \over 
{\lower2pt\hbox{$\sim$}} $}
\newcommand{\Msun}{$M_{\odot}$}

\defcitealias{KilonovaSeekersI}{Killestein and Kelsey et al. \ \citeyear{KilonovaSeekersI}}

\begin{document}

   \title{GOTO065054+593624: a 8.5 mag amplitude dwarf nova identified in real time via Kilonova Seekers}
   \titlerunning{GOTO0650: a high amplitude dwarf nova}

   \author{
   T.~L. Killestein\thanks{ \email{tom.killestein@gmail.com}}\inst{1,2}
   \and G. Ramsay\inst{3}
   \and M. Kennedy\inst{4}
   \and L. Kelsey\inst{5}
    \and D. Steeghs\inst{2}
    \and S. Littlefair\inst{6}
    \and B. Godson\inst{2}
    \and J. Lyman\inst{2}
    \and M. Pursiainen\inst{2}
    \and B. Warwick\inst{2}
    \and C. Krawczyk\inst{7}
    \and L.~K. Nuttall\inst{7}
    \and E. Wickens\inst{7}
    \and S.~D. Alexandrov\inst{8,9}
    \and C.~M. da Silva\inst{8,10,11}
    \and R. Leadbeater\inst{12}
    \and K. Ackley\inst{2}
    \and M.~J. Dyer\inst{6}
    \and F. Jim\'enez-Ibarra\inst{13}
    \and K. Ulaczyk\inst{2}
    \and D.~K. Galloway\inst{13}
    \and V.~S. Dhillon\inst{6,14}
    \and P. O'Brien\inst{15}
    \and K. Noysena\inst{16}
    \and R. Kotak\inst{1}
    \and R.~P. Breton\inst{17}
    \and E. Pallé\inst{14}
    \and D. Pollacco\inst{2}
    \and A. Kumar\inst{2,18}
    \and D. O'Neill\inst{2}
    \and T. Butterley\inst{19}
    \and R. Wilson\inst{19}
    \and S. Mattila\inst{1,20}
    \and A. Sahu\inst{2}
    \and R. Starling\inst{15}
    \and C.~Y. Wang\inst{21}
    \and Q. Liu\inst{22} 
    \and A. Li\inst{23,24}
    \and Z. Dai\inst{25,26}
    \and H. Feng\inst{27}
    \and W. Yuan\inst{28,29}
    \and R. Billington\inst{8}
    \and A.~G. Bull\inst{8,30,31}
    \and S. Gaudenzi\inst{8,10}
    \and V. Gonano\inst{8}
    \and H. Krawczyk\inst{8}
    \and M.~T. Mazzucato\inst{8,31,32}
    \and A. Pasqua\inst{8}
    \and J.~A. da Silva Campos\inst{8,33}
    \and M. Torres-Guerrero\inst{8}
    \and N.~N. Antonov\inst{34}
    \and S.~J. Bean\inst{12}
    \and E.T. Boeneker\inst{10}
    \and S. M. Brincat\inst{10}
    \and G.~S. Darlington\inst{12,10}
    \and F. Dubois\inst{35,36}
    \and F.-J. Hambsch\inst{36,10,37,38}
    \and D. Messier\inst{10,39}
    \and A. Oksanen\inst{10,40}
    \and G. Poyner\inst{12}
    \and F. D. Romanov\inst{10, 41}
    \and I.~D. Sharp\inst{12}
    \and T. Tordai\inst{42}
    \and T. Vanmunster\inst{10,39}
    \and K. Wenzel\inst{38}
          }

    \authorrunning{Killestein et al.}

   \institute{
    Department of Physics \& Astronomy, University of Turku, Vesilinnantie 5, Turku, FI-20014, Finland. 
   \and Department of Physics, University of Warwick, Gibbet Hill Road, Coventry CV4 7AL, UK.
   \and Armagh Observatory \& Planetarium, College Hill, Armagh, BT61 9DG, UK 
   \and School of Physics, Kane Building, University College Cork, Cork, Ireland. 
   \and Institute of Astronomy and Kavli Institute for Cosmology, University of Cambridge, Madingley Road, Cambridge CB3 0HA, UK.
   \and Astrophysics Research Cluster, School of Mathematical and Physical Sciences, University of Sheffield, Sheffield, S3 7RH, UK.
   \and Institute of Cosmology and Gravitation, University of Portsmouth, Portsmouth, PO1 3FX, UK.
   \and Kilonova Seekers c/o Zooniverse, Department of Physics, University of Oxford, Denys Wilkinson Building, Keble Road, Oxford, OX1 3RH, UK.
   \and Institute of Plant Physiology and Genetics, Bulgarian Academy of Sciences, Acad. G. Bontchev Str., Bl.23, Sofia 1113, Bulgaria.
   \and American Association of Variable Star Observers, 185 Alewife Brook Parkway, Suite 410, Cambridge, MA 02138, USA.
   \and Variable Stars South, Royal Astronomical Society of New Zealand, P O Box 3181, Wellington, New Zealand.
   \and British Astronomical Association Variable Star Section, Burlington House, Piccadilly, London, W1J 0DU, UK. 
   \and School of Physics \& Astronomy, Monash University, Clayton VIC 3800, Australia.
   \and Instituto de Astrofísica de Canarias, E-38205 La Laguna, Tenerife, Spain.
   \and School of Physics \& Astronomy, University of Leicester, University Road, Leicester LE1 7RH, UK.
   \and National Astronomical Research Institute of Thailand, 260 Moo 4, T. Donkaew, A. Maerim, Chiangmai, 50180 Thailand.
   \and Jodrell Bank Centre for Astrophysics, Department of Physics and Astronomy, The University of Manchester, Manchester, M13 9PL, UK.
   \and Department of Physics, Royal Holloway - University of London, Egham Hill, Egham, TW20 0EX, UK.
   \and Centre for Advanced Instrumentation, University of Durham, DH1 3LE Durham, UK. 
   \and School of Sciences, European University Cyprus, Diogenes street, Engomi, 1516 Nicosia, Cyprus. 
   \and Department of Astronomy, Tsinghua University, Beijing 100084, China. 
   \and Physics Department, Tsinghua University, Beijing 100084, China. 
   \and School of Physics and Astronomy, Beijing Normal University, Beijing 100875, China; 
   \and Institute for Frontier in Astronomy and Astrophysics, Beijing Normal University, Beijing 102206, China. 
   \and Yunnan Observatories, Chinese Academy of Sciences, Kunming 650216, China. 
   \and Key Laboratory for the Structure and Evolution of Celestial Objects, Chinese Academy of Sciences, Kunming 650216, China. 
   \and Key Laboratory of Particle Astrophysics, Institute of High Energy Physics, Chinese Academy of Sciences, 100049, Beijing, China. 
   \and National Astronomical Observatories, Chinese Academy of Sciences, Beijing 100101, China. 
   \and School of Astronomy and Space Science, University of Chinese Academy of Sciences, Beijing 100049, China. 
   \and Department of Earth and Environmental Sciences, The University of Manchester, Williamson Building, Oxford Road, Manchester, M13 9PL, UK.
   \and Royal Astronomical Society, Burlington House, Piccadilly, London, W1J 0BQ, UK. 
   \and Physical Sciences Group, Siena Academy of Sciences, Piazzetta Silvio Gigli 2, 53100 Siena, Italy. 
   \and Astronomical Soc. of Southern Africa, c/o SAAO POBox 9, Observatory 7935 CT Rep South Africa. 
   \and Institute for Advanced Physical Studies, 111 \say{Tsarigradsko shose} Blvd, Sofia 1784, Bulgaria. 
    \and AstroLAB IRIS, Provinciaal Domein \say{De Palingbeek}, Verbrandemolenstraat 5, 8902 Zillebeke, Ieper, Belgium 
    \and Vereniging Voor Sterrenkunde (VVS), Oostmeers 122 C, 8000 Brugge, Belgium 
    \and Groupe Européen d’Observations Stellaires (GEOS), 23 Parc de Levesville, 28300 Bailleau l’Evêque, France 
    \and Bundesdeutsche Arbeitsgemeinschaft für Veränderliche Sterne e.V. (BAV), Munsterdamm 90, 12169 Berlin, Germany. 
    \and Center for Backyard Astrophysics, New York, NY, USA. 
    \and Hankasalmi Observatory, Hankasalmi, Finland 
    \and Burke-Gaffney Observatory, Saint Mary’s University, 923 Robie Street, Halifax, NS B3H 3C3, Canada. 
    \and Polaris Observatory, Hungarian Astronomical Association, Laborc utca 2/c, 1037 Budapest, Hungary.
   }
   \date{Received xxx; accepted yyy}

 
   \abstract{
    Dwarf novae are astrophysical laboratories for probing the nature of accretion, binary mass transfer, and binary evolution -- yet their diverse observational characteristics continue to challenge our theoretical understanding. We here present the discovery of, and subsequent observing campaign on GOTO065054+593624 (hereafter GOTO0650), a dwarf nova of the WZ Sge type, discovered in real-time by citizen scientists via the Kilonova Seekers citizen science project, which has an outburst amplitude of 8.5 mag. An extensive dataset charts the photometric and spectroscopic evolution of this object, covering the 2024 superoutburst. GOTO0650 shows an absence of visible emission lines during the high state, strong H and barely-detected He~II emission, and high-amplitude echo outbursts with a rapidly decreasing timescale. The comprehensive dataset presented here marks GOTO0650 as a candidate period bouncer, and highlights the important contribution that citizen scientists can make to the study of Galactic transients.
   }


   \maketitle
%
\section{Introduction}
Cataclysmic variables (CVs) are the most populous type of accreting
binary, and represent the end state of binary star evolution in which
the more massive stellar component is a white dwarf (WD) and the
secondary is a late-type main sequence star, or a brown dwarf. At some
point in the past, their WD progenitor star evolved off the main sequence, and
its expanding outer envelope engulfed a main sequence companion
star. During a short-lived common-envelope episode, angular momentum
was removed efficiently from the binary with the ejection of the
envelope, leaving a close pair with a separation of a few solar
radii and orbital period of $\sim$6--10 hours~\citep{Warner2003}.

Gradually, angular momentum is lost from the system through a combination of the stellar wind from the secondary star and gravitational wave radiation, which results in the secondary star filling its Roche lobe and initiating ballistic mass transfer to
the WD through the inner Lagrange point L1, forming an accretion disc
around the WD if its magnetic field is \ltae 1\,MG. Low luminosity states correspond to intervals of low
accretion rate, where accreting matter accumulates in the outer
disk. When the disk density reaches a certain threshold, the disk
becomes ionized and switches viscosity state. The ensuing outburst
manifests as a high accretion rate, high luminosity episode and causes
the `cataclysmic' phenomenon. Indeed, it is these high-amplitude
outbursts which enabled the first detection of a CV in 1855
\citep{Hind1856}. Since then, amateur astronomers (perhaps better known
today as citizen scientists) have continued to make an important
contribution to their study \citep[e.g.][]{Jensen1995,Shears2018}.

Angular momentum continues to be lost from the system until the
late-type star becomes fully convective and magnetic braking is thought to be significantly reduced. The secondary detaches from its Roche lobe and
accretion ceases until the continued angular momentum loss through gravitational
waves causes the binary orbit to shrink, and accretion restarts. This
results in the so-called period gap (see \citealt{Schreiber2024} for a
recent summary). As the CVs orbit shrinks further, a point (known as the 'period minimum', predicted to be $\sim$78 min) is reached when the core of the late-type star
becomes a brown dwarf. At this point, further mass loss from the donor then causes
its radius to stop contracting, which to conserve angular momentum, the orbital period starts to increase, and thus the CV becomes a \say{period bouncer}, i.e., a CV system which has evolved beyond the period minimum (see \citealt{Knigge2011} for a review of the evolution of CVs). 

The number of CVs near the predicted period minimum is a
sensitive test of binary evolution models, which predict that a
significant fraction (40-70\%) of all galactic CVs are period bouncers \citep[e.g. ][]{Kolb1993,Howell2001,GoliaschNelson2015,Belloni2020}: the problem is that very few such systems are known {\citep[e.g. ][]{Pala2018,Inight2023}}, possibly because for systems around the period minimum it is difficult to determine whether they have still to reach the minimum, or whether their period is now getting longer and the system is therefore post-minimum.

One class of CV with orbital periods close to the period minimum is the \object{WZ Sge} type (see \citealt{Kato2015} for a review), which are good
candidates for period bouncers. WZ Sge systems show high amplitude outbursts (typically greater than 6 mag) which recur on a time scale of years or decades -- much longer than dwarf nova outbursts in systems with longer orbital periods -- due to their very low mass transfer rate \citep[e.g. $10^{15}$g/s ($1.6\times10^{-10}$ \Msun/yr) for WZ Sge stars compared to $10^{16}$g/s ($1.6\times10^{-9}$ \Msun/yr) for SU UMa stars;][]{OsakiMeyer2002}. 
They also display a periodic modulation in the long decline from maximum, known as superhumps, with a period slightly longer than the orbital period. Some outbursts show evidence of `echo' outbursts which occur 2-3 weeks after the peak and are excellent tests of accretion instability models \citep[e.g.][]{Patterson2002}.

In recent years, optical wide-field surveys have been discovering more WZ Sge systems than ever before~\citep[e,g. ][]{Tampo2021}, filling out the parameter space of such systems, and unveiling unique examples. In this paper, we present observations and initial analysis of a new WZ Sge system with a number of intriguing properties, discovered using the GOTO all-sky survey, and first identified via the Kilonova Seekers citizen science project in Oct 2024.  

\section{Discovery}
GOTO065054+593625 (hereafter GOTO0650) was first detected during the all-sky survey of the Gravitational-wave Optical Transient Observer (GOTO; \citep{Dyer2018,Steeghs2022, Dyer2024}) in an image taken on 2024 Oct 04 03:36:36 UT by GOTO-North on La Palma. The source was extremely bright on the discovery, with a GOTO $L$ band (400-700 nm) magnitude of $13.37\pm0.01$, located at \(\alpha = 06^\text{h} 50^\text{m} 54.49^\text{s}\), \(\delta = +59^\circ 36' 24.51''\) (J2000)
~\citep{GOTO0650DiscoveryATel}. The field was last visited $\approx2$\,d prior to discovery by GOTO, with a non-detection of this source down to $L<20.2$ mag. There was a further non-detection 5.5\,h later with the All-Sky Automated Survey for Supernovae~\citep[ASAS-SN; ][]{Shappee2014}, $g>17.4$ mag, constraining the outburst within a time window of 1.8 days.\footnote{All times in this paper are given relative to the GOTO discovery epoch (MJD 60,587.15) -- likely close to the time of maximum light given the strong rise constraint from GOTO survey photometry.}

\begin{figure}
    \centering
    \includegraphics[width=\linewidth]{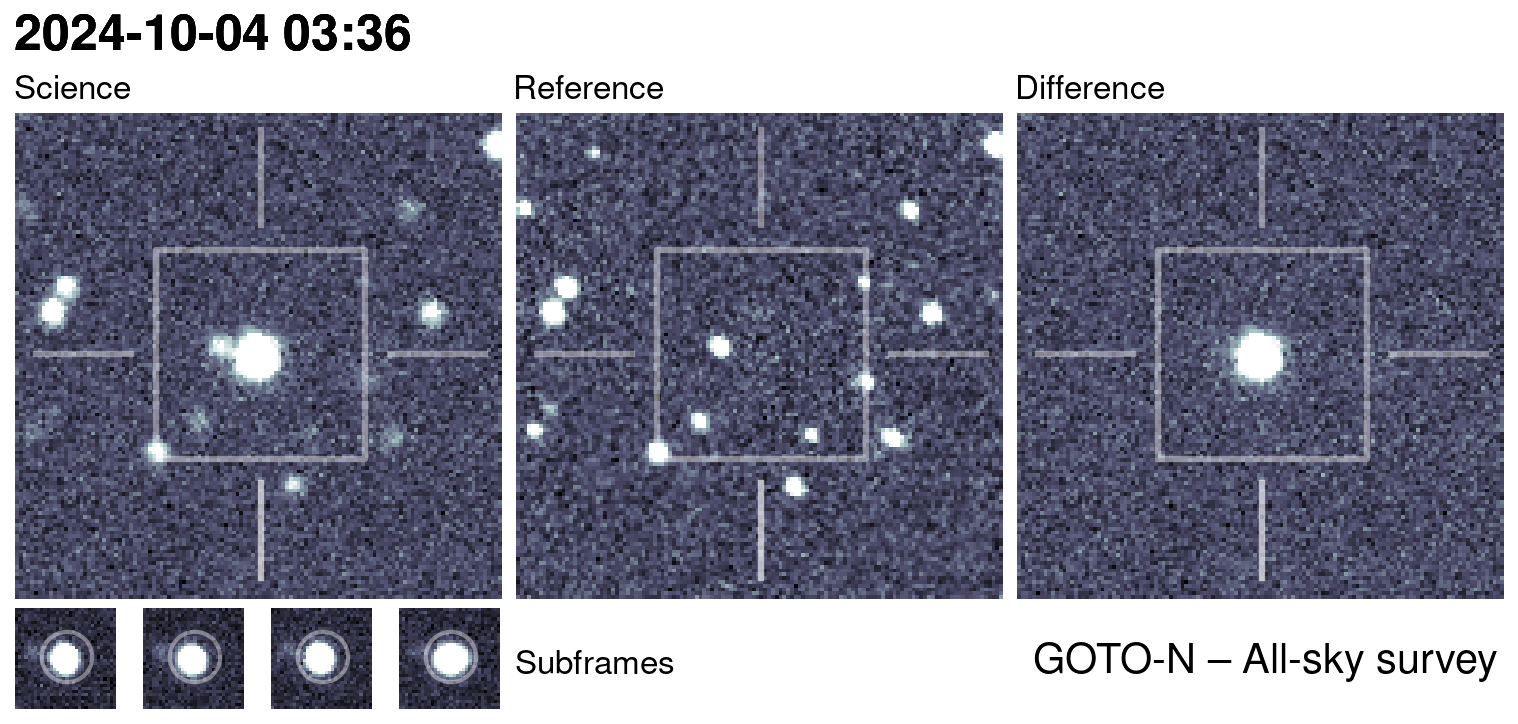}
    \caption{GOTO discovery triplet of GOTO0650 as seen by citizen scientists on the Kilonova Seekers project, with science frame, historical reference image, and difference image. The inner box corresponds to a scale of $\approx80$ arcsec across. The labelling indicates the time of observation of this source, and the GOTO site and survey mode that the observation was taken from. Below the science image are a series of subframes showing the four individual exposures that are stacked to create the science image.}
    \label{fig:discovery-figure}
\end{figure}

Notably, GOTO0650 was first identified by citizen scientists working via the Kilonova Seekers~\citepalias{KilonovaSeekersI} project on the Zooniverse platform. The volunteers reached the required threshold to promote the object as real approximately 3.5 hours after discovery, with 90\% of volunteers who viewed the candidate image voting the object as real and astrophysical. Given the primarily extragalactic focus of GOTO team vetting, this object had previously been put on hold in the \say{pending} source list of the internal GOTO Marshall (Lyman et al., in prep.), awaiting further information given its possible association to a faint point source identified in the Panoramic Survey Telescope and Rapid Response System (Pan-STARRS) DR1 imaging~\citep{Chambers2016}. Without the Kilonova Seekers volunteers flagging this object, rapid follow-up would not have been possible, and this object may have been missed entirely. Citizen science is a powerful methodology for driving discoveries in vast datasets that must be sifted in a focused way by scientists, leaving ample room for novel serendipitous discoveries. The discovery image of GOTO0650 is presented in Figure~\ref{fig:discovery-figure}, as seen by the volunteers.
There is a clear underlying blue source at this position in DESI Legacy Survey~\citep{Dey2019} and Pan-STARRS 4$\pi$~\citep{Chambers2016} Survey imaging, which we identify as GOTO0650 in quiescence. The source has magnitudes $g=21.84$, $r=21.68$ and $z=22.02$ in Legacy Surveys DR10, with ($g-r\approx0.2$). Taken with the GOTO discovery magnitude, this implies an outburst of $\approx8.5$ mag in amplitude.

We inspected available forced photometry (photometric observations extracted at the known position) of GOTO0650 in quiescence from GOTO, the Asteroid Terrestrial-impact Last Alert System~\citep[ATLAS; ][]{Tonry2018}, ASAS-SN, and the Zwicky Transient Facility~\citep[][]{Bellm2019} via the GOTO Lightcurve Service~(Jarvis et al., in prep.), ATLAS Forced Photometry Server~\citep{Shingles2021}, ASAS-SN Sky Patrol~\citep{Kochanek2017}, and ZTF Forced Photometry Service~\citep{Masci2023} respectively, and find no evidence for prior outbursts during the timespan covered (approx. 4 years). We engaged in a more comprehensive search for historical outbursts using plate archives, and discuss a candidate outburst in 1951 in more detail in Section~\ref{sec:observations}.

Follow-up spectroscopy with the Nordic Optical Telescope (NOT) and Liverpool Telescope \citep[LT; ][]{Steele2004} obtained $\approx$ 4 days post-discovery confirmed the Galactic nature of this source~\citep{GOTO0650ClassificationATel}, showing a blue continuum with Balmer absorption lines at $z=0$, weak He~I $\lambda4471$, and weak Na~ID absorption. The spectrum
showed no emission lines present at the time of classification, consistent with observations of the WZ Sge system \object{GW Lib} at maximum light \citep{Hiroi2009}. Given the changing nature of the spectrum of GW Lib (and other systems) over its outburst, we triggered an intensive observing campaign, to investigate the nature of the accretion disc over the outburst duration.

\section{Observations and data reduction}
\label{sec:observations}
This section presents the detailed photometric and spectroscopic observations, along with the data reduction process, for GOTO0650. Upon discovery, we promptly initiated an intensive follow-up campaign utilising multiple facilities. The first follow-up photometry of GOTO0650 was captured just 30 minutes after the Kilonova Seekers platform issued the alert, and the first spectrum was obtained within 4 days.

\subsection{Photometry}
Upon discovery, we quickly began an intensive photometric follow-up campaign of GOTO0650, using a number of facilities (summarised in Table~\ref{tab:overview_phot}). The first imaging of GOTO0650 was taken just 30 minutes after the Kilonova Seekers platform alerted us to this discovery.
\begin{table}
    \centering
    \caption{Summary of all photometric data obtained on GOTO0650 in this work.}
    \begin{tabular}{l l l}
    \hline
     Facility & Bands & Number of images\\
     \hline
     GOTO & $L$ & 52 \\
    LCO 0.4m & $ugriz$ & 326 \\
    \textit{Swift} & $UVW1$,$UVM2$,$UVW2$,$U$ & 12 \\
    TTT & $ugriz$ & 78 \\
    pt5m & $BVRI$ & 72 \\
    LJT & $CV$ & 1 \\
    \hline
    \end{tabular}
    \label{tab:overview_phot}
\end{table}

We obtained a sequence of $ugriz$ photometry with the 0.4m robotic telescopes of the Las Cumbres Observatory Global Telescope Network \citep[LCOGT;][]{Brown2013} through the Kilonova Seekers - LCO: STAR (Surveying Transients with Amateur Researchers) program\footnote{\url{https://lco.global/education/partners/kilonova-seekers/}}, a LCO Global Sky Partner. The Kilonova Seekers - LCO: STAR educational program was specifically designed as an extension of the GOTO Kilonova Seekers citizen science project to enable volunteers to be involved in all aspects of transient astronomy: from discovery, to triggering further observations, to classification; learning how to reduce data, and how to make and analyse light-curves and colour images with guidance and support from the Kilonova Seekers team. 
A mixture of high-cadence observations and staged follow-up were performed. Initial observations consisted of a continuous sequence of 30s exposures for a 30-minute window in $g$-band to identify any short-term variability, with daily follow-up in $ugriz$ bands to characterise the decline of the light curve. Data reduction was performed with the BANZAI pipeline \citep{McCully2018}.

Nightly observations were obtained with the pt5m 0.5m telescope~\citep{Hardy2015} at Roque de los Muchachos, La Palma. Observations were obtained with $BVRI$ filters, and were calibrated and stacked with custom pipelines.

Observations were also obtained with the 80 cm TTT1 telescope, part of the Two-meter Twin Telescopes\footnote{\url{https://ttt.iac.es}} project, in the $ugriz$ filters with a iKon-L 936 CCD. Data were calibrated with a custom pipeline. Observations were also obtained on 2024 Dec 3 using the Lijiang 2.4m telescope with a 2 min exposure in the clear (CV) band.

Photometry on all reduced frames was performed with a custom \texttt{photutils}-based \citep{Bradley2016} seeing-matched aperture photometry pipeline (Killestein et al., in prep.) to ensure all photometry was obtained in a standardised way. All $griz$ photometry is calibrated to PanSTARRS-1 \citep{Chambers2016} field stars, while $u$-band photometry is calibrated to synthetic photometry derived from the \textit{Gaia} DR3 BP-RP spectra of stars in the field~\citep{Montegriffo2023}. Johnson-Cousins $BVRI$ photometry is approximately calibrated to PanSTARRS using synthetic magnitudes for field stars derived using the transformations in \citet{Tonry2012}. The LJT data were corrected to $V$-band using source colour information estimated from LCO and TTT data.
All photometry is corrected for a foreground reddening of $E_{B-V}=0.056$ using the 3D reddening maps of \citep{Lallement2022,Vergely2022}, evaluated at the inferred distance of GOTO0650 (see Section~\ref{sec:results}), assuming a \citet{Fitzpatrick1999} extinction curve when adjusting our spectra. We keep this reddening estimate fixed throughout, but propagate uncertainty arising from the distance forwards in parameter estimates.
Magnitudes are given in their natural systems (AB for $ugriz$, Vega for $BVRI$). The collated photometry of GOTO0650 is presented in Figure~\ref{fig:lightcurve}, and summarised in Table~\ref{tab:photom}.

We also include all available photometry from GOTO survey observations in our light curve. GOTO images are processed with the \texttt{kadmilos} pipeline~(Lyman et al., in prep.), and forced photometry at the location of GOTO0650 performed using the GOTO Lightcurve service~(Jarvis et al., in prep.).

We obtained two epochs (2024 Oct 19, 2024 Oct 20) of high-cadence photometry with \textit{gotito}, the GOTO CMOS Pathfinder instrument sited next to the GOTO-N node (see Godson et al., in prep.). Observations were predominantly taken in the $L$-band filter at 30~s cadence using a Sony IMX411 CMOS sensor. Low-level image calibration was performed with custom pipelines. For inclusion on light-curve plots and tables, we calibrate \textit{gotito} to $g$-band, accounting for the colour term using the observed $g-r$ colour derived from LCO observations close in time.

Photometry from a number of citizen scientist observers was retrieved via the American Association of Variable Star Observers (AAVSO) International Database\footnote{\url{https://www.aavso.org/aavso-international-database-aid}}, in both Johnson-Cousins ($BVR$) and $CV$ bands -- although only $V$ and $CV$ observations are included in this work. To remove offsets and standardise these data, we compute per-observer offsets based on overlapping intervals of LCO/TTT $gri$ data (transformed to Johnson $V$ with the relations of \citealt{Tonry2012}). We include only observations in this standardisation process where $V<17$ mag, to ensure robust corrections are obtained. We interpolate this \say{reference} light curve to each observer's individual photometric epochs with a thin plate spline\footnote{\url{https://github.com/treverhines/RBF}}, and compute the (sigma-clipped) median deviation across the observers' photometry as the offset. Per-observer shifts with few exceptions are small, typically $<0.1$\,mag. Nevertheless, correcting for these is crucial to ensure smooth light curves across the poorly-sampled regions in our multi-colour photometry, such as the dip phases, where intrinsic scatter is higher.

\begin{figure}
    \centering
    \includegraphics[width=\linewidth]{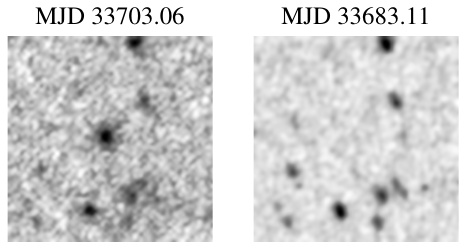}
    \caption{Harvard photographic plates retrieved via DASCH depicting a historical GOTO0650 superoutburst on MJD 33,3703.06 (1951.15), after correcting the alignment for additional distortion. We also show an image taken on MJD 33683.11 (1951.037).}
    \label{fig:plates}
\end{figure}

Furthermore, we searched for evidence of historical outbursts using the lightcurve tool provided by the Digital Access to a Sky Century @ Harvard (DASCH) project \citep{Grindlay2012}. A single source was recovered at a distance of $13''$ from the position of GOTO0650 with a detection on 1951 Feb 26 (MJD 33,703.06) at an apparent magnitude of 13.0 (Plate ac41904) with a non-detection 20 days prior (Plate rh15570). We show cutouts of these plates in Figure~\ref{fig:plates}
Whilst this initial separation between our measured position of GOTO0650 and this putative historical superoutburst is large (c.f. the plate scale of 4$''$\,pix$^{-1}$), there is some evidence for unaccounted-for distortions in the plate local to GOTO0650. Recomputing the astrometry with a smaller cut-out of the plate (via \url{astrometry.net};~\citealt{Lang2010}) yields better concordance ($\approx2$ pixels). We therefore take this source detection to be an outburst from GOTO0650, and elaborate further on the implications of this historical superoutburst in Section~\ref{sec:discussion}.

\subsection{Spectroscopy}

We obtained a number of spectra with the Alhambra Faint Object Spectrograph (ALFOSC) mounted on the NOT at Roque de los Muchachos Observatory, La Palma. Specifically, grisms \#4, \#7, and \#17 were used. These spectra were reduced with the PyNOT\footnote{\url{https://jkrogager.github.io/pynot/}} pipeline.
Further spectra were obtained with the SPectrograph for the Rapid Acquisition of Transients (SPRAT; \citealt{Piascik2014}) on the LT~\citealt{Steele2004}), and reduced with Pypeit~\citep{Prochaska2020}.
Spectra were also obtained with an ALPY  spectrograph using 200/600 l/mm grisms (45/12Å resolution) mounted on a 0.28m telescope at Three Hills Observatory. The spectra were reduced using ISIS\footnote{\url{http://www.astrosurf.com/buil/isis-software.html}}
and calibrated in relative flux using reference stars at matching airmass. These spectra were retrieved via the BAA Spectroscopy Database~\footnote{\url{https://britastro.org/specdb}}
All spectra were rescaled with a polynomial correction factor to match contemporaneous photometry, to ensure any subtle changes in continuum shape are robust and to ensure good absolute flux calibration.
The full details of all spectroscopic observations included in the paper are tabulated in Table \ref{spec_table}.


\begin{table*}[h!]
\caption{Log of spectra obtained of GOTO0650 during our observing campaign where Phase indicates the time since outburst.}
\label{spec_table}
\centering
\begin{tabular}{lllllllll}
\hline
Instrument & Grism & Date & Phase & Exp. time & Airmass & Slit width & $R$ & Range\\
    & & (UT) & (d) & (s) &     & (arcsec) & & (Å)\\
\hline
NOT/ALFOSC & 4 & 2024 Oct 08 05:00:30 & 4.1 & 120 & 1.24 & 1.0 & 360 & 3200-9600\\
LT/SPRAT & Blue & 2024 Oct 08 05:25:57 & 4.1 & 200 & 1.21 & 1.8 & 350 & 4020–8100\\
THO/ALPY200 & & 2024 Oct 11 00:50:32 & 6.9 & 1201 & 1.27 & 3.1 & 166 & 3750-7500 \\
LT/SPRAT & Blue & 2024 Oct 12 04:30:55 & 8.0 & 240 & 1.26 & 1.8 & 350 & 4020–8100\\
THO/ALPY200 & & 2024 Oct 13 02:01:40 & 8.9 & 3005 & 1.14 & 3.1 & 166 & 3750-7500 \\
THO/ALPY200 & & 2024 Oct 15 00:26:12 & 10.9 & 3606 & 1.29 & 3.1 & 166 & 3750-7500 \\
LT/SPRAT & Blue & 2024 Oct 17 03:26:06 & 13.0 & 300 & 1.36 & 1.8 & 350 & 4020–8100\\
NOT/ALFOSC & 7 & 2024 Oct 20 02:38:31 & 16.0 & 300 & 1.47 & 1.0 & 650 & 3650-7110\\
LT/SPRAT & Blue & 2024 Oct 24 03:51:48 & 20.0 & 300 & 1.25 & 1.8 & 350 & 4020–8100\\
THO/ALPY600 & & 2024 Oct 27 00:41:45 & 22.9 & 7215 & 1.18 & 3.1 & 518 & 3750-7500 \\
NOT/ALFOSC & 7 & 2024 Nov 01 05:49:44 & 28.1 & 3600 & 1.16 & 1.0 & 650 & 3650-7110\\
NOT/ALFOSC & 7 & 2024 Nov 10 01:38:17 & 36.9 & 1800 & 1.43 & 1.0 & 650 & 3650-7110\\
\hline
\end{tabular}
\end{table*}

\subsection{X-ray \& UV observations}
\label{obs_xray}

Observations were made using the Follow-up X-ray telescope (FXT) on
the Einstein Probe \citep{EP2022}. The FXT has a passband between 0.5-10 keV and a spatial resolution of 20-24$^{''}$ (half-power diameter) on axis. The first of seven observations
was made 2.48 d after the discovery with the most recent being made 34.67 d after discovery. A significant, if weak, X-ray source was detected at a position consistent with the optical position. 

Observations of GOTO0650 were also made by the Neil Gehrels \textit{Swift}
satellite, with seven epochs in total (PIs: Ramsay, Bhattacharaya, Chandra) with the start time being 3.9 to 34.1 d after discovery. The observations using the X-ray Telescope (XRT) were generally short ($<$1 ks) and thus an X-ray source was not detected at the position of GOTO0650 during any of the pointings.

Simultaneous UV observations with the Ultra-Violet/Optical Telescope (UVOT) \citep{Roming2005} detected GOTO0650 in all pointings. Observations were scheduled in \say{filter of the day} mode, with GOTO0650 being observed in the $UVW2$, $UVM2$, $UVW1$ and $U$ bands respectively. Given GOTO0650 is a point source and concretely detected in all pointings, we use the standard UVOT source catalogs. The UV flux of GOTO0650 broadly traces the optical evolution, as expected -- we elaborate further on the detailed evolution in latter sections.

\section{Analysis and results}
\label{sec:results}

\subsection{Photometric evolution}

\begin{figure*}
    \centering
    \includegraphics[width=\linewidth]{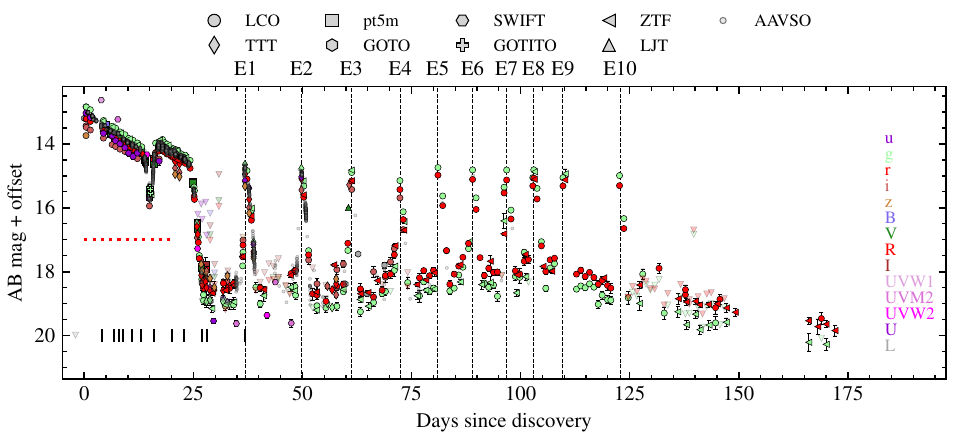}
    \caption{
     Collated photometry of GOTO0650 (see Table~\ref{tab:overview_phot}), shown relative to the GOTO discovery epoch. An overview of this photometry and the associated uncertainties are presented in Table~\ref{tab:photom}. Error bars are typically too small to be seen. Photometry has been corrected for Galactic extinction. Vertical markers indicate phases at which spectra were obtained. Non-detections are overplotted with the triangular markers where available, corresponding to 5$\sigma$ upper limits - with opacity for ease of visualisation. The vertical lines annotated $E_N$ correspond to the N$^\mathrm{th}$ echo outburst observed, with the red dotted line indicating the magnitude at which we distinguish between high and low states.
    }
    \label{fig:lightcurve}
\end{figure*}

The photometric series of GOTO0650 is comprehensive, with dense coverage through the first 2 months of the light curve evolution. In Figure~\ref{fig:lightcurve}, we present the photometric evolution of GOTO0650, both from our intensive monitoring, curated survey data, and AAVSO observers. There are a number of key phases of the light curve of this object that we focus on in this section -- initial discovery, the decline from the peak/plateau phase, an early dip, and then a deeper decline into the pre-quiescent phase, showing repeated strong outbursts.

The initial superoutburst of GOTO0650 was detected in an image from the GOTO all-sky survey obtained 2024 Oct 4 03:36:36 at $L=13.36$ mag, uncorrected for extinction. The exact outburst date is constrained by a prior GOTO observation (2024 October 2 03:36:32) which reached a 5-sigma limiting magnitude of $L=20.14$. This implies a rise rate of $\gtrsim3.4$ mag d$^{-1}$. In the absence of any more constraining non-detections, we take the time of outburst to be MJD $60,586.2\pm1.0$ - the midpoint of the first detection and last non-detection, with an uncertainty of half the span.
The true outburst date likely lies in the second half of this interval, based on rise rates ($\approx7$ mag/day) from uninterrupted coverage of other WZ~Sge-like systems with Kepler~\citep[e.g.][]{Ridden-Harper2019}. At peak, the $V$-band apparent magnitude of the light curve is $m_V = 13.04 \pm 0.05$, corrected for foreground extinction as noted in Section~\ref{sec:observations}. We estimate the distance to GOTO0650 by using $M_{V}=4.5\pm0.3$ derived from the peak absolute magnitudes of known WZ Sge systems \citep{Patterson2011}. Taking the peak $V=13.04$ inferred above, we determine a distance of $510^{+80}_{-60}$ pc.

\begin{figure}[b]
    \centering
    \includegraphics[width=\linewidth]{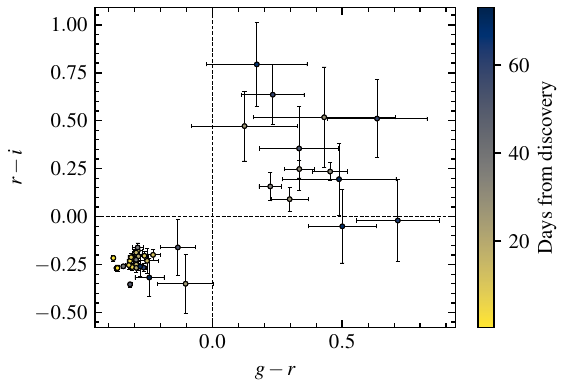}
    \caption{Colour locus plot of GOTO0650 from LCO and TTT $gri$ photometry, showing the distinct two clusters corresponding to the high and low state. Points are shaded to show the phase relative to discovery.}
    \label{fig:colourlocus}
\end{figure}

\begin{figure}[b]
    \centering
    \includegraphics[width=\linewidth]{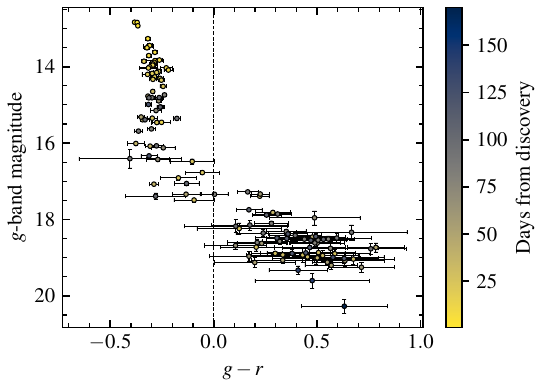}
    \caption{The $g-r$ colour of GOTO0650, as a function of apparent magnitude. The colour scale shows the phase relative to discovery. A clear change in source colour with magnitude is visible showing a strong reddening on the way down to the low state.}
    \label{fig:hysteresis}
\end{figure}

Over the next 14 days, GOTO0650 showed a steady decline of $\approx0.1$ mag d$^{-1}$ in all bands, estimated via linear regression. The colour (see Figure~\ref{fig:colourlocus}) over this time also showed a marginal evolution, reddening in both $g-r$ and $g-i$ colour indices by $\approx0.1$ mag. Figure~\ref{fig:hysteresis} shows the overall co-evolution of the colour indices with respect to the outburst phase.
Spectroscopy during this phase showed a blue continuum with strong Balmer absorption (discussed in greater detail in Section~\ref{sec:spectra}), indicative of the emission being dominated by a geometrically thick disk.

\begin{figure}
    \centering
    \includegraphics[width=\linewidth]{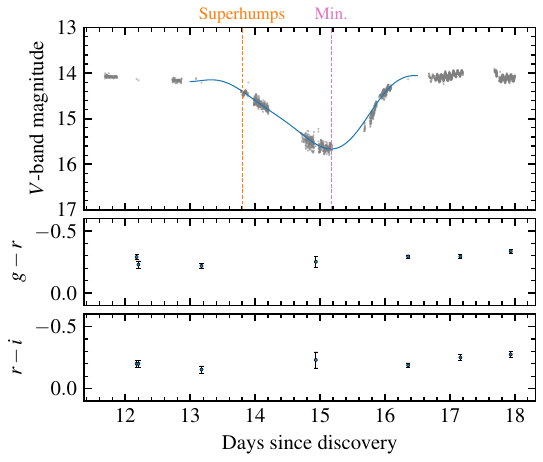}
    \caption{The high-cadence $V$-band photometry around the dip observed in GOTO0650 from AAVSO, LCO, pt5m, and TTT. The blue curve is a spline interpolation of the data. The vertical dashed line shows the initial onset of superhumps as the dip began. The $g-r$ and $r-i$ colour indices show no significant evolution through the dip. The dip shows a marked asymmetry, with a slow ingress and rapid egress. High-amplitude superhumps of $\approx0.1$mag amplitude are visible immediately after the dip.}
    \label{fig:dip}
\end{figure}

At +14\,d post-discovery, GOTO0650 entered a short and shallow dip in brightness, similar to those seen in WZ Sge systems~\citep{Kato2015}. A zoom-in is depicted in Figure~\ref{fig:dip}. The depth is $\approx1.5$ mag in $V/R$-band, and lasts for only 4 days - with a slightly skewed profile, with a slower drop in ($t_{1/2}=0.58$ d) and more rapid rise out ($t_{1/2}=0.87$ d). The dip is essentially achromatic, with no significant change in colour indices, as evident from our photometry obtained at the bottom of the dip. Approximately around this time (+13.8\,d, see Section~\ref{sec:shorttimescalevar}), ordinary superhumps emerged, showing a characteristic asymmetric profile, contributing to excess scatter in the lower-cadence light curve. This is consistent with the timing of the ignition of superhumps in EG Cnc \citep{Patterson1998}, which we note has a light curve very similar to that of GOTO0650. After the conclusion of this short-lived event, GOTO0650 rises to $\approx0.1$ mag brighter than on the previous plateau, before declining at a very similar rate.

The main dimming event in GOTO0650 began +25\,d post-discovery, dropping from $g=15$ mag into a persistent, low state of around $g=19$ mag. The decline is rapid, at $\approx1.5$mag d$^{-1}$ in $V$-band. The transition to the low state shows a stark shift in colour -- transitioning from $g-r\approx-0.3$ to $g-r\approx0.5$ mag, with a similar evolution mirrored in $r-i$ (Figure \ref{fig:hysteresis}).

\subsection{Echo outbursts}

Approximately 36.5 days after the initial superoutburst, GOTO0650 entered a phase punctuated by a number of \say{echo outbursts}, short-duration rebrightenings which have been observed in a relatively small number of CVs \citep[e.g. ][]{Patterson1998,Kato2004}. GOTO0650 has shown ten rebrightenings, E1-10 hereafter, which reach close to the pre-dip brightness (typically with an amplitude of $\approx3.7$ mag), and appear blue in colour. The sampling of the photometry covering echo bursts E3-10 is lower than during E1-2, but there is some hint that the durations of bursts E3-10 are shorter than E1-2.

\begin{figure}
    \centering
    \includegraphics[width=\linewidth]{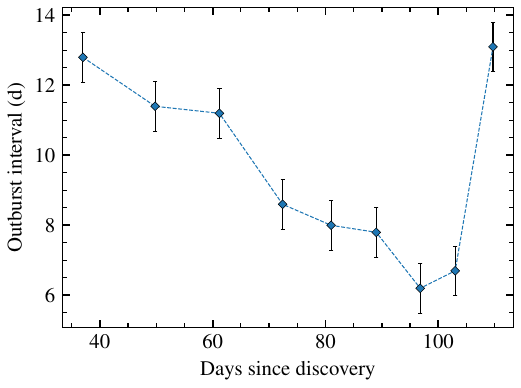}
    \caption{The inter-burst timescale for the echo outbursts observed in GOTO0650, with error bars corresponding to the half-width of a typical outburst. GOTO0650 shows a marked evolution towards more frequent echo outbursts.}
    \label{fig:echotimescale}
\end{figure}

We note the time of the echo outbursts and the time since the previous outburst in Table \ref{tab:echo-times} and Figure \ref{fig:echotimescale}. Between E2 and E9 the recurrence time decreases from 13.5 d to 6.5, with an increase between E10 and E9 of 16.0 d. The mean recurrence time is 8.7 d and 7.9 d if we exclude E10. This follows the same trends seen in EG Cnc which showed six echo outbursts in its 1996 outburst with a mean recurrence time of 7.1 d \citep{Patterson1998}. The mean time between echo outbursts in GOTO0650 are longer than most WZ Sge systems reported in \citet{Kato2015} and \citet{Meyer2015} which are generally 3-5 d. There is no colour change seen in $g-r$ accompanying this change in outburst state. Although the sampling precludes a definitive statement, there is a hint that the brightness level between outbursts increases as the outburst proceeds (see \citet{Meyer2015}.

\begin{table}
\begin{tabular}{lrr}
\hline
Burst number & Time since outburst & Time since     \\
           &    (d)            & previous burst (d) \\
\hline
E1 & 36.5 & -- \\
E2 & 50.0 & 13.5 \\
E3 & 61.0 & 9.0 \\
E4 & 72.5 & 11.5 \\
E5 & 81.0 & 8.5 \\
E6 & 89.0 & 8.0 \\
E7 & 96.0 & 7.0 \\
E8 & 103.0 & 7.0 \\
E9 & 109.5 & 6.5 \\
E10 & 125.5 & 16.0 \\
\hline
\end{tabular}
\caption{The time of the echo outbursts (see Figure \ref{fig:lightcurve}) since the discovery date and
the time delay since the previous echo outburst.}
\label{tab:echo-times}
\end{table}

No further echo outbursts were observed beyond E10 on day 125 -- with confident detections from both our LCO photometry and ZTF over this time period. Our latest observations, around 160 days post-discovery, show GOTO0650 continues to decline slowly in brightness, returning to quiescence -- likely indicating this is the conclusion of this superoutburst of GOTO0650.

\subsection{Short-timescale variability}
\label{sec:shorttimescalevar}
Throughout the evolution of GOTO0650, high-cadence photometry was obtained from both citizen scientists and professional observatories. This photometry was initially devoid of obvious variability for around the first two weeks of evolution, with no evidence for superhumps emerging during this timeframe at the precision of the available data, primarily from AAVSO observers. The first credible detection of ordinary superhumps is seen in the observations of Vanmunster (reported in \texttt{vsnet-alert 28024})\footnote{\url{http://ooruri.kusastro.kyoto-u.ac.jp/mailman3/hyperkitty/list/vsnet-alert@ooruri.kusastro.kyoto-u.ac.jp/message/WZ7VAQUUGELF22D54Q3YXVGLXN7QTN6Q/}} on 2024 Oct 18, approximately 13.8 days post-discovery. The onset of superhumps is neatly illustrated in Figure~\ref{fig:dip}, with the epoch of onset shown with the orange dashed line. This length of delay in appearance places it on the longer side of the distribution shown by WZ Sge systems (c.f. Fig.~18 of \citealt{Kato2015}) and coincides with the beginning of the first dip discussed in the prior section. Uncertainties on the superhump parameters at this first epoch are large owing to only $\approx$1 cycle being observed, so we do not include this epoch in any analysis. 

\begin{figure}
    \centering
    \includegraphics[width=\linewidth]{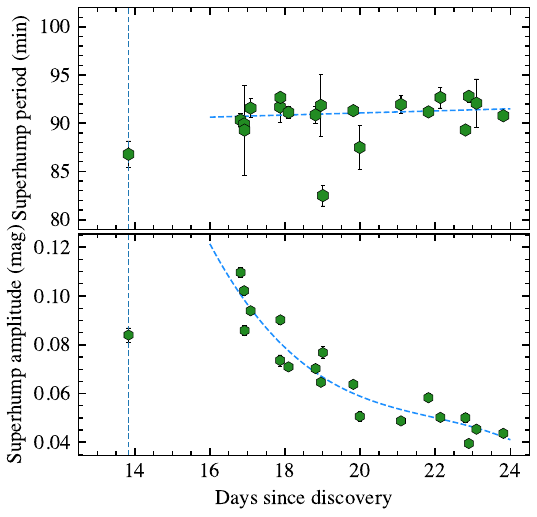}
    \caption{Superhump period and amplitude evolution in GOTO0650, with time given relative to the GOTO discovery date. The blue-dashed line indicates the first epoch at which superhumps are detected. Error bars on both plots are 1$\sigma$ statistical uncertainties. An illustrative fit to both the superhump period and amplitude is overplotted, to show the overall evolution.}
    \label{fig:superhumps}
\end{figure}

\begin{figure}
    \centering
    \includegraphics[width=\linewidth]{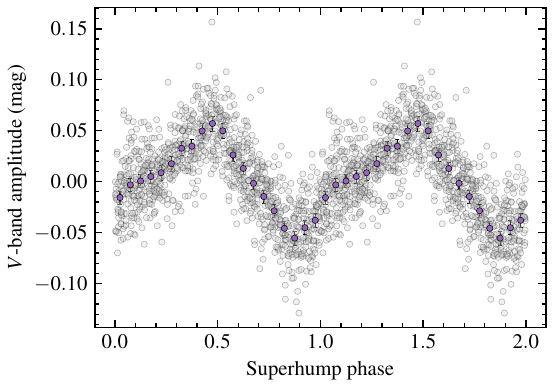}
    \caption{
    Phase-averaged light curves from AAVSO observers between day 16 and 18, revealing the profile of the observed superhumps close to their maximum amplitude.
    The phase is given arbitrarily relative to discovery date, given an imprecise measure of when the superhumps first emerged.}

    \label{fig:superhump-profile}
\end{figure}

To investigate the emergence of superhumps, and evolution of the period, we inspect and compute Lomb-Scargle~\citep{Lomb1976,Scargle1982} periodograms of the individual high-cadence time-series observations in $V$, $CV$, and $g$ bands. Photometry is detrended with a linear fit to remove longer-timescale evolution. Figure~\ref{fig:superhumps} illustrates the superhump period and superhump amplitude as a function of time post-discovery to probe their evolution.
Approximate polynomial fits (linear in superhump period, and cubic in superhump amplitude) are overplotted to guide the eye.
Uncertainties on the best-fitting period are derived from the FWHM of the periodogram peak. This is a purely statistical error and does not fold in potential systematic uncertainties. The superhump amplitude is estimated as the peak-to-peak amplitude of the best-fitting sinusoid, with uncertainties conservatively estimated from the standard error of the residuals. We also phase-fold the AAVSO light curves around the time of maximum superhump amplitude (days 16-18), and combine them together in phase bins of 0.05 to better reveal the superhump profile from the scattered data. Points in each phase bin are combined with the inverse-variance weighted mean, and plotted in Figure~\ref{fig:superhump-profile}. There is a clear asymmetry in the profile.

The superhump period is approximately $91\pm1$\,min around 17\,d post-discovery, and remains broadly consistent with this throughout, showing some marginal evolution to longer periods. Given the uncertainties in the period estimation, it is not trivial to robustly estimate the period derivative $\dot{P}$. Based on the revised stellar evolutionary models of \citet{Knigge2011} the mass ratio, $q=M_{1}/M_{2}$, is 0.05 or 0.12 depending if it is a post- or pre-minimum binary. Assuming the super-humps arise due to a tidal instability at a radius at the 3:1 resonance with the secondary, we use the relationship given by Equation 8 of \cite{KatoOsaki2013}, appropriate to the mean of stage B superhumps, to determine an orbital period of 90.4 or 88.5 min (post/pre-period minimum) with typical statistical uncertainties of 1\,min. 
The superhump amplitude shows a strong evolution -- reaching a maximum around +16\,d post-discovery of 0.1\,mag peak-to-peak, roughly after the first dip concludes, then declines rapidly after this phase before being largely undetectable beyond 24 days. Observations of the 2021 outburst of \object{V627 Peg} show that prior to the first decline phase, the amplitude of the superhumps was only $\sim$0.05 mag \citep{Tampo2023}, less than observed in GOTO0650. 
 A more detailed characterisation of the superhumps, particularly $O-C$ diagrams and similar analyses requiring precise phasing are deferred to a follow-up paper.

\subsection{Spectral evolution}
\label{sec:spectra}

\begin{figure}
    \centering
    \includegraphics[width=\linewidth]{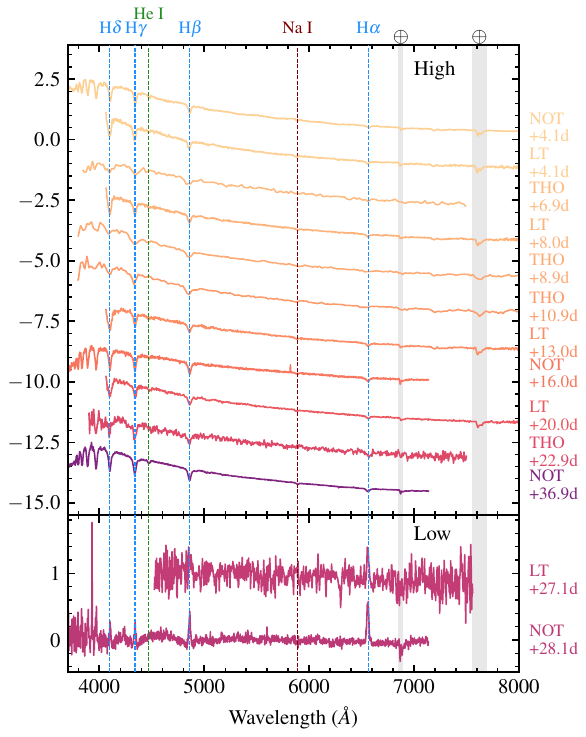}
    \caption{The spectral of GOTO0650 split into high and low state for visualisation purposes. All spectra are flux-calibrated to contemporaneous photometry, and corrected for Galactic extinction. Key spectral features are indicated by dashed lines, and the telescope and epoch of each observation are displayed.}
    \label{fig:spectral-series}
\end{figure}

The spectral evolution of GOTO0650 is shown in Figure~\ref{fig:spectral-series}. Given the clear differences between the two states, we split the spectra into \say{high} and \say{low} states for further analysis. In both states, we compute line parameters (equivalent width, flux, FWHM), which we tabulate in Table~\ref{tab:lineparams}.

\begin{table*}
\centering
\caption{Line parameters (equivalent width, FWHM, and line flux) of various key spectral features seen in the optical spectra of GOTO0650 throughout its' evolution. Uncertainties correspond to 1$\sigma$ confidence intervals, estimated via Monte Carlo resampling. Line fluxes are only given for emission lines, and we preserve the sign of EW to flag emission and absorption.}
\begin{tabular}{l|lll|lll|lll}
\hline
& \multicolumn{3}{c|}{H$\alpha$} & \multicolumn{3}{c|}{H$\beta$} & \multicolumn{3}{c}{H$\gamma$} \\
Phase & EW & FWHM & F$_{\rm H\alpha}/10^{-15}$ & EW & FWHM & F$_{\rm H\beta}/10^{-15}$ & EW & FWHM & F$_{\rm H\gamma}/10^{-15}$ \\
 (d) & (\AA) & (km/s) & (erg/cm$^2$/s) & (\AA) & (km/s) & (erg/cm$^2$/s) & (\AA) & (km/s) & (erg/cm$^2$/s) \\
\hline
4.1 & $6.1\pm0.2$ & 1880 & - &$7.7\pm0.2$ & 1900 & - &$8.1\pm0.2$ & 1920 & -  \\
4.1 & $6.1\pm0.3$ & 1750 & - &$7.3\pm0.3$ & 1810 & - &$9.0\pm0.3$ & 2040 & -  \\
8.0 & $6.7\pm0.3$ & 1460 & - &$8.2\pm0.2$ & 2050 & - &$9.2\pm0.3$ & 2120 & -  \\
13.0 & $6.8\pm0.4$ & 1720 & - &$8.5\pm0.3$ & 2170 & - &$8.9\pm0.4$ & 2070 & -  \\
16.0 & $8.5\pm0.3$ & 1720 & - &$11.7\pm0.2$ & 1870 & - &$12.7\pm0.3$ & 2000 & -  \\
20.0 & $6.8\pm0.4$ & 1650 & - &$9.0\pm0.3$ & 2180 & - &$8.9\pm0.4$ & 2000 & -  \\
28.1 & $-13.6\pm0.6$ & 1010 & $1.9\pm0.1$ & $-9.0\pm0.6$ & 1000 & $1.3\pm0.7$ & $-6.5\pm0.7$ & 760 & $0.9\pm0.1$ \\
36.9 & $7.6\pm0.1$ & 1730 & - &$10.9\pm0.1$ & 1930 & - &$12.1\pm0.1$ & 2080 & -  \\
\hline
\end{tabular}
\label{tab:lineparams}
\end{table*}

The spectra in the high state show a predominantly blue continuum with strong Balmer absorption features. H$\alpha$ appears strongly suppressed compared to the other Balmer lines. This is expected when the accretion disc is optically thick, as seen in \citet{Tovmassian2022}, who show spectra of the WZ Sge system V455 And over its 2007 superoutburst.
There are also weak He and Na absorption lines present in the spectrum, though only well-detected in high-quality spectra.
Our higher-resolution NOT spectra (grism 7, $R\approx650$) at +16\,d, show no evidence of weak emission in line cores.

The absorption features show some evolution in equivalent width through the outburst phase, which matches well with the overall light curve evolution -- with the lines deepening as the continuum drops away. There is a clear increase in the equivalent width of all lines around +15\,d, corresponding to the initial dip, followed by a drop as the overall luminosity increases.

\begin{figure}
    \centering
    \includegraphics[width=\linewidth]{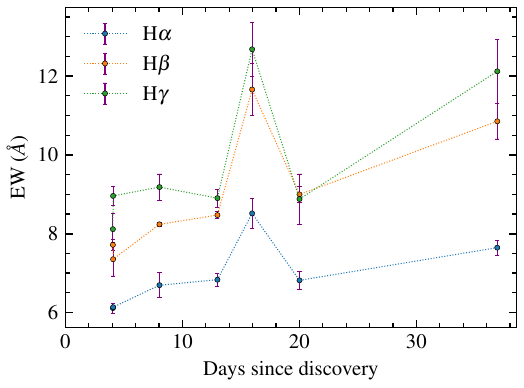}
    \caption{Equivalent widths and uncertainties for the most prominent absorption features in the high state for GOTO0650. Uncertainties are estimated via Monte Carlo resampling of the spectra. The EWs show an overall increase over time, with a marked jump around the time of the first dip.}
    \label{fig:equivwidth}
\end{figure}

\begin{figure}[]
    \centering
    \includegraphics[width=\linewidth]{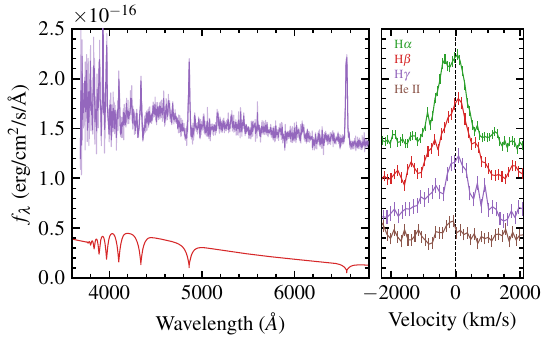}
    \caption{The NOT/ALFOSC gr7 spectrum of GOTO0650 obtained at +28.1\,d, at the bottom of the decline off plateau ($g\approx19$). A 5-pixel variance-weighted boxcar filter has been applied to reduce noise. The inset panel is centred on the prominent H$\alpha$ emission line, showing tentative hints of a more complex profile, but ultimately limited by spectral resolution. The WD contribution is estimated from the best-fitting model SED estimated in Section~\ref{SEDquiet}, scaled to the distance inferred in Section~\ref{sec:results} assuming a fixed $\log g$ of 8, and a radius inferred from the sequences of \citet{Bedard2020}.}
    \label{fig:quiescentzoom}
\end{figure}

We obtained observations in the low state on days 27 and 28 respectively, with LT/SPRAT and NOT/ALFOSC. Despite the poor signal-to-noise of the SPRAT spectrum, there is a clear detection of H$\alpha$ in emission. The NOT/ALFOSC spectrum the next day shows a wealth of spectral features -- narrow Balmer emission lines clearly detected up to H$10$, very weak He~II$\lambda4686$, Fe~II$\lambda5168$, strong Ca~II H\&K emission, and some broad absorptions superimposed on a flat continuum.  The line widths of emission features are $\approx1000$ km s$^{-1}$, and only marginally resolved in our spectrum. This spectrum is depicted in Figure~\ref{fig:quiescentzoom}, alongside a WD spectrum approximately matching GOTO0650 in quiescence, rescaled by a factor of 10 to enhance visibility.

In contrast to WZ Sge and V455 And ~\citep[e.g.][]{Gilliland1986,Tovmassian2022}, where complex He~I$\lambda$5876,6676 features are present, at a level that would be easily detected in our NOT/ALFOSC spectrum, the spectrum of GOTO0650 lacks these features. However, other WZ Sge systems such as V386 Ser \citep[e.g.][]{Inight2023} do not show strong He~I and He~II emission. Considering H$\alpha$, there are clear hints of a double-peaked emission profile arising from the accretion disk, although it is not well resolved. This may be due to the low spectral resolution ($R\approx650$), or alternatively, it may suggest a low binary inclination \citep{HorneMarsh1986}. All Balmer lines present show a pronounced asymmetry between the red and blue peaks -- with a slightly enhanced red component. Although measuring the precise separation between the two maxima is difficult, the velocity separation of the two profile maxima in H$\alpha$ is $\approx370$\,km s$^{-1}$.
The ratios of EWs of H$\alpha$:H$\beta$:H$\gamma$ in the low state are 2.2:1.4:1, and the equivalent widths themselves are broadly lower than typical cataclysmic variables in quiescence \citep[e.g. ][]{Szkody2004}. 

\begin{figure}[b]
    \centering
    \includegraphics[width=\linewidth]{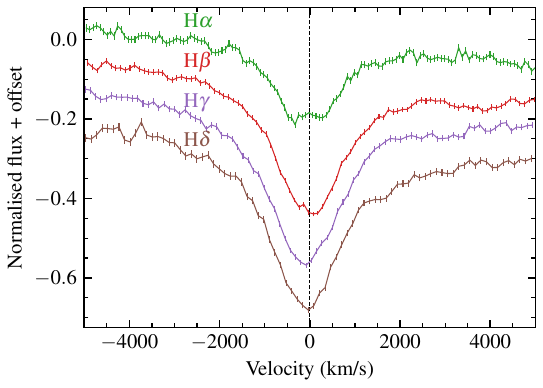}
    \caption{The Balmer lines as observed in the +36\,d spectrum obtained at the peak of the first echo outburst. There are clear differences between the line profiles moving to higher-ionisation lines, with the varying line profiles well-explained by the differences in amplitude between the Balmer emission seen in the low-state spectrum obtained on 28\,d.}
    \label{fig:outburst_spec}
\end{figure}

Some limited spectroscopy was obtained during the echo outburst phases, but was largely hampered owing to their short timescale. The spectrum obtained on day 36, at the peak of the first echo outburst, shows a H$\alpha$ profile with a double dip where the line core appears to be filled in by emission, with more typical profiles seen in the other lines. Figure~\ref{fig:outburst_spec} illustrates these line profiles, normalised to the continuum. 


Further spectroscopy deeper into quiescence is required to constrain system parameters like the orbital period $P$ and binary mass ratio $q$, which are challenging to constrain robustly with the dataset presented here.

\subsection{Spectral energy distribution in quiescence}
\label{SEDquiet}

\begin{figure}
    \centering
    \includegraphics[width=\linewidth]{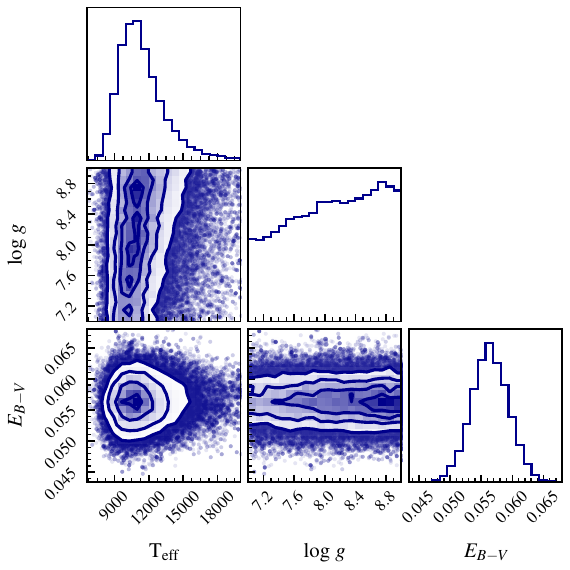}
    \caption{Posterior distributions from modelling of the available PanSTARRS photometry of GOTO0650 in quiescence. The reddening parameter is essentially prior-dominated, but a reasonable constraint on effective temperature is obtained.}
    \label{fig:wd_constraints}
\end{figure}

There is some limited archival photometry of GOTO0650 in the quiescent state in the PanSTARRS (\citealt{Chambers2016}; $grizy$ bands) and DESI Legacy Survey (\citealt{Dey2019}; $grz$ bands). Although we lack detections in bands bluer than $g$ -- most notably lacking NUV photometry to tightly constrain the temperature, and NIR to better know the nature of the donor star -- there is still some utility in modelling the SED to attempt to infer parameters of the system. This is useful to consider the evolutionary implications for the WD in the GOTO0650 system. 
Under the assumed distance modulus of $\mu=8.3\pm0.2$ derived from the peak absolute magnitude of WZ Sge systems~\citep{Patterson2011}, GOTO0650 has an $V$-band absolute magnitude in quiescence of $M_V = 13.1\pm0.2$ -- compatible with absolute magnitudes for both WDs and the donor stars in these types of systems. Given the blue observed colour, it is a safe assumption that the majority of this flux arises from the WD in the system. 
We proceed under this assumption and do not see any strong excess in the redder bands nor donor spectral features, though note that GOTO0650 is not formally detected in the $y$-band.
No astrometric solution for GOTO0650 is available from \textit{Gaia} as of the current data releases, making more detailed characterisation of the system challenging in absence of a parallax.
To constrain the properties of the WD in GOTO0650, we infer the temperature and surface gravity using the PanSTARRS observed colours in conjunction with synthetic photometry derived from the Bergeron\footnote{\url{https://www.astro.umontreal.ca/~bergeron/CoolingModels/}} model grids~\citep{Holberg2006, Tremblay2011}, and the \citet{Bedard2020} evolutionary sequences assuming a thick H envelope. 
Evolutionary sequences were interpolated using the \texttt{MR\_relation}\footnote{\url{https://github.com/mahollands/MR_relation}} Python code
Given the poorly constrained distance, derived from the empirical \citealt{Patterson2011} relation rather than geometric parallax, we use only colour information in our fitting. 
We consider only the available PanSTARRS photometry in this analysis, to avoid any potential systematic errors associated with mixing survey magnitude systems.
We assume a prior uniform over the whole $\mathrm{T_{eff}}$-$\log g$ grid, and assuming a fixed reddening value, though propagate uncertainties (primarily stemming from distance estimate) to avoid underestimating uncertainties in the WD parameters. We assume the observed colours $g-r$, $r-i$, $i-z$, and $z-y$ can be modelled with a product-Gaussian likelihood, and sample the posterior distribution using \texttt{emcee}~\citep{Foreman-Mackey2013} package. The sampler was run for 3000 steps using 32 walkers, with the first 225 steps in each chain (5$\times$ the integrated autocorrelation length, $\tau$; \citealt{Goodman2010}) discarded as burn-in. The resultant posterior distributions are shown in Figure~\ref{fig:wd_constraints}.
From the marginalised parameter distributions, we derive $\mathrm{T_{eff}} = 11100^{+2200}_{-1500}$ K for the WD, accounting for the uncertainties in reddening and an unknown $\log g$. As expected, poor constraints on $\log g$ are obtained from photometry alone, with spectroscopy in the quiescent phase being most useful to determine the value. Our derived temperature (11,000\,K) for the WD in GOTO0650 is towards the lower end of the effective temperature distribution of WDs in CV systems~\citep[e.g. ][]{Pala2017}, and is directly comparable with a number of period bounce candidates, e.g. \object{QZ Lib}~\citep[10,500K; ][]{Pala2018}, \object{EZ~Lyn}~\citep[12,000K; ][]{Aviles2010}. Stronger constraints on the effective temperature, via NUV photometry to constrain the Balmer jump, or deep spectroscopy to probe the WD spectral features, will be crucial to cement this clue towards the nature of GOTO0650. We here assume that we are indeed seeing the WD, when components from the disk and accretion column may be present.

\subsection{X-ray fluxes and luminosity}

\begin{figure}
    \centering
    \includegraphics[width=\linewidth]{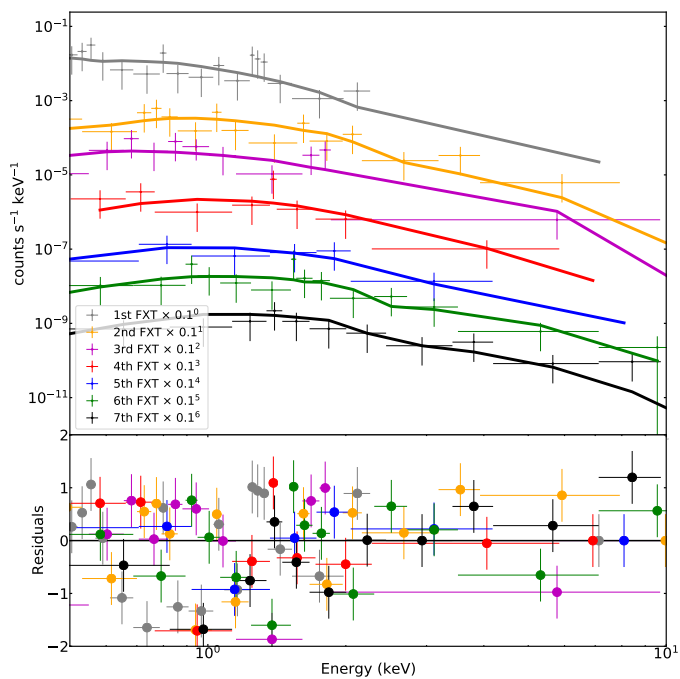}
    \caption{The fits and residuals to the seven X-ray spectra taken using the EP FXT instrument using an absorbed power law.}
    \label{fig:xrayfits}
\end{figure}

In Section~\ref{obs_xray} we outline the observations made using the EP and {\sl Swift} X-ray telescopes. Using standard analysis tools, we determined the observed flux in the 0.5--10 keV band using an absorbed power-law. We show the unabsorbed flux; photon index and luminosity over time are shown in Table~\ref{xrayflux} and the spectral fits in Figure \ref{fig:xrayfits}. There is evidence that the X-ray flux was highest shortly after maximum optical brightness and again between 30--35 d after discovery during the first main dip phase.

\begin{table}
\caption{The unabsorbed fluxes, photon index and luminosity of GOTO0650 determined using EP FXT data
obtained at seven epochs where we indicate the time since the outburst. The
luminosity is derived assuming a distance of 510 pc.}
\begin{tabular}{llll}
\hline
Time & Flux & Photon Index & Luminosity\\
(days) & (erg s$^{-1}$ cm$^{-2}$) & & (erg s$^{-1}$) \\
\hline
2.48 & 10.6$^{+2.0}_{-1.1} \times 10^{-14}$ & 3.23$\pm$0.29 & 3.3$^{+0.6}_{-0.3}\times10^{30}$ \\
6.02 & 7.8$^{+2.4}_{-1.0} \times 10^{-14}$ & 2.04$\pm$0.32 & 2.4$^{+0.8}_{-0.3}\times10^{30}$ \\
10.52 & 8.2$^{+3.7}_{-2.5} \times 10^{-14}$ & 1.92$\pm$0.33 & 2.6$^{+1.1}_{-0.7}\times10^{30}$ \\
14.88 & 8.0$^{+2.9}_{-1.1} \times 10^{-14}$ & 1.62$\pm$0.32 & 2.5$^{+0.9}_{-0.3}\times10^{30}$ \\
30.04 & 4.2$^{+2.9}_{-1.7} \times 10^{-14}$ & 1.50$\pm$0.49 & 1.3$^{+0.9}_{-0.5}\times10^{30}$ \\
30.56 & 11.0$^{+4.5}_{-3.3} \times 10^{-14}$ & 1.11$\pm$0.30 & 3.4$^{+1.4}_{-1.0}\times10^{30}$ \\
34.67 & 12.3 $^{+3.4}_{-2.7} \times 10^{-14}$ & 1.10$\pm$0.22 & 
3.8$^{+1.1}_{-0.8}\times10^{30}$ \\
\hline
\end{tabular}
\label{xrayflux}
\end{table}

Assuming a distance of 510 pc, the X-ray luminosity of GOTO0650 is in the range 1.3--3.8$\times10^{30}$ erg s$^{-1}$. The
slope of the power law was significantly steeper during the first epoch. 
X-ray observations of GW Lib and SSS J122222-311525 show a decline in X-ray flux from maximum optical light \citep{Neustroev2018}. The X-ray luminosity of GOTO0650 is an order of magnitude lower than that of SSS J1222.

We merged the seven epochs of {\sl Swift} XRT data to obtain an event file with a combined exposure time of 6.96 ksec. From this event file, we made an image
using {\tt xselect}\footnote{\url{http://heasarc.nasa.gov/docs/software/lheasoft/ftools/xselect/xselect.html}}. We
used the HEASoft tool {\tt XIMAGE} and the routine {\tt SOSTA} (which
takes into account effects such as vignetting, exposure and the point
spread function) to determine a 3$\sigma$ upper limit at the position of GOTO0650 of 2.65$\times10^{-3}$ ct/s. Using {\tt PIMMS}\footnote{\url{https://heasarc.gsfc.nasa.gov/cgi-bin/Tools/w3pimms/w3pimms.pl}} we assume a hydrogen column density of 3.9$\times10^{20}$ cm$^{-2}$ (consistent with $A_{V}=0.22$) and determine an unabsorbed flux (0.5--10 keV) of 9.7$\times10^{-14}$ for a photon index of 1.8 (the mean of that derived using EP data). 
The emission detected by Einstein Probe is often below the obtained upper limits from the summed \textit{Swift} XRT observations. 
There is no evidence for historical X-ray emission from GOTO0650 in a number of high-energy surveys. ROSAT All-Sky Survey \citep[RASS; ][]{Voges1999} observations in 1990 yield an upper limit of $2.2\times10^{-13}$  erg s$^{-1}$ cm$^{-2}$ in the 0.2-2\,keV band.

\section{Discussion}
\label{sec:discussion}

With an outburst amplitude of $\sim$8.5 mag, a likely orbital period of
$\sim$90 min and ten echo outbursts,
GOTO0650 shares all the characteristics of an WZ~Sge type accreting binary. The light curve overall appears similar to the 1996/1997 outburst of EG Cnc \citep{Patterson1998}. The recurrence time of the echo outbursts in GOTO0650 becomes shorter, with a decreasing amplitude with a slow increase in the inter-outburst quiescent brightness.
The 1951 superoutburst identified in DASCH plates (see Figure~\ref{fig:plates}) 
indicates there was one outburst 73 years ago. Although we cannot discount that outbursts have been missed in these intervening years, it does indicate a likely outburst recurrence time of many years or decades. WZ Sge and EG Cnc show outbursts every few decades.

\subsection{Super-humps}

One key observational property of WZ Sge binaries is the presence of
super-humps. However, at this stage, it is unclear what observations
were made in the first week of the outburst and how sensitive they were to any
super-hump feature. The available AAVSO photometric coverage shows no significant periodicity prior to +13.8\,d, but do not robustly rule out low-amplitude periodicities. If super-humps were indeed not present, the fact they
were only detected two weeks after the outburst is unusual: for instance AL
Com showed super-humps within the first week of an outburst in 1995
\citep{Patterson1996}. However, \citet{OsakiMeyer2002} showed that the
absence of early super-humps could be due to the low inclination of
the system -- potentially reinforced by the low-EW, narrowly-separated double-peak profile seen in our quiescent spectrum.

\subsection{Temperature of the white dwarf}

Because of their long recurrence time between outbursts, WZ Sge
systems are excellent objects to determine the temperature of the WD
in the absence of accretion events and to determine the cooling time
rate (years) over the post outburst \citep[e.g. ][]{Godon2006}. We know
the temperature of the white dwarf in GOTO0650 in quiescence is approximately
 10-13 kK (Section~\ref{SEDquiet}) but is likely to be 2--3 times
hotter during the immediate post-peak epoch. Although the temperature
is not especially well constrained, the WD could lie beyond the cool
edge of the WD instability strip \citep[e.g.][]{VanGrootel2012}. It is
therefore possible that GOTO0650 could pass through the WD instability
strip as it cools. Observations of WZ Sge showed photometric
pulsations around 28\,s as it declined from outburst
\citet[e.g. ][]{Welsh2003}. Since the period was not stable, it was not a signature of the WD rotation period. It remains unclear whether
its origin is g-mode pulsations (low-frequency gravity-mode pulsations due to horizontal displacements of stellar gas) of the WD or due to channelled magnetic accretion onto the WD. Interestingly, \citet{CastroSegura2025} obtained multi-band high-cadence photometry of GOTO0650 during the decline from echo outburst, identifying a periodicity of $148.5\pm0.4$\,s  which is markedly longer than what was seen for WZ Sge's pulsations. \citet{CastroSegura2025} outline possible origins for their observed  periodicity.   

\subsection{Space density of CVs}

Although there are many thousands of known or suspected CVs, there
remains a tension between their predicted space density and the number
which are actually known. The study of \citet{Pala2020}, which used
Gaia data, identified all the known CVs out to a distance of 150
pc. The observed fraction of period bounce CVs in this sample is less
by at least a factor of 5 compared to the predicted number. The super-hump period and the narrow emission lines seen, implying weak mass transfer, point to GOTO0650 being a candidate period bounce CV. Confirmation will need phase resolved spectroscopy before the system has returned to quiescence, as it
would then be too faint ($g=21.8$), making this a crucial component to understanding GOTO0650.

\subsection{Citizen science}

Even in the age of widespread wide-field imaging surveys, the discovery
of galactic transient events with amplitudes $>$8 mag is relatively rare. 
Since they can be identified quickly, and are likely to take some considerable time to return to quiescence, they provide a great opportunity to perform multi-wavelength observations over the duration of the event to study phenomena such as echo outbursts, which in turn give insight to how accretion discs can repeatedly go from stable to unstable states. 

Such rapid
requests for observations are not always possible on large telescopes. However, citizen scientists are often able to spend
entire nights observing a single source at very short notice, generating high-quality data. A geographically-distributed ensemble of such scientists can determine (as demonstrated by Figure~\ref{fig:superhumps}) how the super-hump period evolves over the outburst, which can then be used to understand conditions in the accretion disc and test theoretical models.

The second key aspect for followup observations is spectroscopy. As
demonstrated in our case, citizen scientists are able to obtain low-resolution spectroscopy, even of relatively faint systems, which is more than adequate to reveal whether specific lines are in absorption or emission. However, 2m-class telescopes are generally required for higher-resolution spectroscopy of fainter targets such as GOTO650, which can reveal weak emission lines in absorption cores. Even higher resolution ($R\gtrsim2000$) spectra from larger telescopes are required to search for radial velocity variations, which
can constrain the masses of the binary components, and measure the orbital period -- pinpointing the evolutionary pathway of short-period binaries. Observations such as these are essential if we are to determine whether objects such as GOTO0650 are indeed period bouncer systems, having evolved through the period minimum.

\section{Conclusions}

In this paper, we present the discovery of a very high amplitude
binary system, GOTO0650, which we have revealed to be a WZ Sge system. This system was discovered, observed, and characterised by citizen scientists -- further underscoring the vital contribution to variable star science that they routinely make. Our intensive followup observations have revealed an outburst light curve similar to the `king of the echo outbursts' EG Cnc \citep{Patterson1998}, 
with superhumps only being detected two weeks after the outburst. The spectra of GOTO0650 have shown only one instance where emission lines were detected, showing only in quiescence strong H emission, with very weak/non-detected He emission. The ten detected echo outbursts of GOTO0650 have shown a number of intriguing characteristics -- rapidly decreasing recurrence timescales, decreasing amplitude and duration, and an overall brightness increase in the low state.
Continued photometric and spectroscopic observations of GOTO0650, particularly high-resolution time-resolved spectroscopy, are required before the system returns to quiescence -- to constrain the radial velocities of the binary components and thus determine if GOTO0650 is indeed a period bouncer, and better understand the nature of this system.

\begin{acknowledgements}
We thank the anonymous referee for their insightful comments, that greatly improved the manuscript.

TLK acknowledges support from the Turku University Foundation (grant no. 081810).
TLK acknowledges a Warwick Astrophysics prize
postdoctoral fellowship made possible thanks to a generous philanthropic donation.

LK acknowledges support for an Early Career Fellowship from the Leverhulme Trust through grant ECF-2024-054 and the Isaac Newton Trust through grant 24.08(w).

JL, MP, and DON acknowledge support from a UK Research and Innovation Fellowship (MR/T020784/1).

BW and BG acknowledge the UKRI’s STFC studentship grant funding, project reference ST/X508871/1.

AK is supported by the UK Science and Technology Facilities Council (STFC) Consolidated grant ST/V000853/1.

RK acknowledges support via the Research Council of Finland (grant 340613).

SM acknowledges support from the Research Council of Finland project 350458.

AS acknowledges the Warwick Astrophysics PhD prize scholarship made possible thanks to a generous philanthropic donation.

RW and TB acknowledge financial support from Science and Technology Facilities Council (STFC, grant number ST/X001075/1). The pt5m telescope is supported by the Isaac Newton Group of Telescopes in La Palma.

RS acknowledges support from the Leverhulme Trust grant RPG-2023-240.

Armagh Observatory and Planetarium is core funded by the Northern Ireland Executive through the Dept of Communities.

The Gravitational-wave Optical Transient Observer (GOTO) project acknowledges the support of the Monash-Warwick Alliance; University of Warwick; Monash University; University of Sheffield; University of Leicester; Armagh Observatory \& Planetarium; the National Astronomical Research Institute of Thailand (NARIT); Instituto de Astrofísica de Canarias (IAC); University of Portsmouth; University of Turku. We acknowledge support from the Science and Technology Facilities Council (STFC, grant numbers ST/T007184/1, ST/T003103/1, ST/T000406/1, ST/X001121/1 and ST/Z000165/1).

This transient was discovered with the assistance of the GOTO Kilonova Seekers Citizen Scientists on Zooniverse. This publication uses data generated via the Zooniverse.org platform, development of which is funded by generous support, including a Global Impact Award from Google, and by a grant from the Alfred P. Sloan Foundation.

We acknowledge with thanks the variable star observations from the AAVSO International Database contributed by observers worldwide and used in this research.

This work has made use of observations made by the Las Cumbres Observatory network of 0.4m telescopes, as part of the LCOGT Global Sky Partners project \say{Kilonova Seekers - LCO: STAR} (PIs: L. Kelsey and T. Killestein). The authors thank E. Gomez for his support through the Global Sky Partners program.

The Liverpool Telescope is operated on the island of La Palma by Liverpool John Moores University in the Spanish Observatorio del Roque de los Muchachos of the Instituto de Astrofisica de Canarias with financial support from the UK Science and Technology Facilities Council.

Based on observations made with the Nordic Optical Telescope, owned in collaboration by the University of Turku and Aarhus University, and operated jointly by Aarhus University, the University of Turku and the University of Oslo, representing Denmark, Finland and Norway, the University of Iceland and Stockholm University at the Observatorio del Roque de los Muchachos, La Palma, Spain, of the Instituto de Astrofisica de Canarias.
The data presented here were obtained in part with ALFOSC, which is provided by the Instituto de Astrofisica de Andalucia (IAA) under a joint agreement with the University of Copenhagen and NOT. 

This article includes observations made in the Two-meter Twin Telescope (TTT) sited at the Teide Observatory of the Instituto de Astrofísica de Canarias (IAC), that Light Bridges operates in the Island of Tenerife, Canary Islands (Spain). The Observing Time Rights (DTO) used for this research were provided by Light Bridges, SL.

This work is based on the data obtained with Einstein Probe, a space mission supported by Strategic Priority Program on Space Science of Chinese Academy of Sciences, in collaboration with ESA, MPE and CNES (Grant No. XDA15310000).
We acknowledge the support of the staff of the Lijiang 2.4m telescope. Funding for the telescope has been provided by Chinese Academy of Sciences and the People's Government of Yunnan Province
Z.-B.D. acknowledges support from the CAS Light of West China Program, the Yunnan Youth Talent Project, the Yunnan Fundamental Research Projects (grant No. 2016FB007, No. 202201AT070180).

This work has made use of data provided by Digital Access to a Sky Century @ Harvard (DASCH), which has been partially supported by NSF grants AST-0407380, AST-0909073, and AST-1313370. Work on DASCH Data Release 7 received support from the Smithsonian American Women’s History Initiative Pool.

This research has used data, tools or materials developed as part of the EXPLORE project that has received funding from the European Union’s Horizon 2020 research and innovation programme under grant agreement No 101004214.
\end{acknowledgements}

\bibliographystyle{aa}
\bibliography{references}
\newpage
\begin{appendix}
\onecolumn
\section{Tables and Additional Figures}
\begin{table*}[h!]
    \centering
    \caption{All obtained photometry of GOTO0650 presented in this work. Magnitudes are uncorrected for Galactic extinction. A machine-readable version of this table is available online.}
    \begin{tabular}{l l l l l l}
    \hline
    Date & Phase & Filter & Magnitude & Error & Telescope \\
    (MJD) & (d) & & (AB)  \\
    \hline 
    60587.15 & +0.0 & L & 13.37 & 0.00 & GOTO \\
    60587.55 & +0.4 & u & 13.31 & 0.02 & LCO 0.4m \\
    60587.55 & +0.4 & r & 13.37 & 0.01 & LCO 0.4m \\
    60587.55 & +0.4 & i & 13.55 & 0.01 & LCO 0.4m \\
    60587.55 & +0.4 & z & 13.83 & 0.04 & LCO 0.4m \\
       &      &   &  \vdots     &      &     \\
    60756.09 & +168.9 & r & 19.62 & 0.18 & LCO 0.4m \\
    \hline
    \end{tabular}
    \label{tab:photom}
\end{table*}
\end{appendix}

\end{document}